\documentclass[12pt]{article}
\usepackage{jheppub}
\usepackage{mathtools}
\usepackage{enumerate}
\usepackage{bm}
\usepackage{dsfont}
\usepackage{blkarray}

\def\para{\paragraph{}}
\def\be#1\ee{\begin{align}#1\end{align}}
\def\nn{\nonumber}
\def\eqn{\eqref}

\newcommand{\tr}{\mathrm{Tr}}
\newcommand{\vol}{\mathrm{Vol}}
\newcommand{\ket}[1]{|{#1}\rangle}
\newcommand{\LambdaR}{\Lambda[\mathcal{R}]}
\newcommand{\DeltaR}{\Delta[\mathcal{R}]}
\newcommand{\mathscale}[2]{\scalebox{#1}{\ensuremath{\displaystyle{#2}}}}

\title{What Symmetries are Preserved by a Fermion Boundary State?}

\author{Philip Boyle Smith}
\author{and David Tong}

\affiliation{
Department of Applied Mathematics and Theoretical Physics, \\
University of Cambridge, Cambridge, CB3 OWA, UK
}

\emailAdd{pb594@damtp.cam.ac.uk}
\emailAdd{d.tong@damtp.cam.ac.uk}

\abstract{Usually, a left-moving fermion in $d=1+1$ dimensions reflects off a boundary to become a right-moving fermion. This means that, while overall fermion parity $(-1)^{F}$ is conserved, chiral fermion parity for left- and right-movers individually is not. Remarkably, there are boundary conditions that do preserve chiral fermion parity, but only when the number of Majorana fermions is a multiple of 8. In this paper we classify all such boundary states for $2N$ Majorana fermions when a $U(1)^N$ symmetry is also preserved. The fact that chiral-parity-preserving boundary conditions only exist when $2N$ is divisible by 8 translates to an interesting property of charge lattices. We also classify the enhanced continuous symmetry preserved by such boundary states: the state with the maximum such symmetry is the $SO(8)$ boundary state, first constructed by Maldacena and Ludwig to describe the scattering of fermions off a monopole.}

\begin{document}

\setcounter{tocdepth}{2}
\maketitle

\section{Introduction}

What symmetries survive when fermions are given a mass?

\para
Within perturbative field theory, the answer is obvious: adding a fermion bilinear to the action breaks chiral symmetries in even spacetime dimension, and some combination of parity and flavour symmetries in odd spacetime dimension. Outside of the perturbative regime, the answer is much less obvious. It is plausible that fermions can be gapped preserving any symmetry that is not protected by an anomaly. This phenomenon sometimes goes by the name of \emph{symmetric mass generation}.

\para
Finding a dynamical mechanism for symmetric mass generation is not straightforward. Indeed, it is only well-understood in low dimensions. The paradigmatic example is in $d=1+1$ dimensions, where a collection of 8 Majorana fermions can be gapped while preserving a $\mathbb{Z}_2 \times \mathbb{Z}_2$ chiral symmetry, corresponding to left- and right-moving fermion parity \cite{fk,ryugso,qi,ryu2}. (See \cite{bentov,carl} for reviews.) When the $d=1+1$ fermions are viewed as edge modes, this calculation underlies the $\mathbb{Z}_8$ classification of interacting SPT phases in $d=2+1$.

\para
The same physics can be seen if the system is placed on a manifold with boundary. Indeed, as emphasised in \cite{shin1,shin2}, there is an intimate connection between gapped phases, and boundary conditions for conformal field theories. Intuitively, one could envisage turning on a gap in one half of space. Any low energy excitation, incident from the gapless side, would view the gapped region as a boundary. Moreover, if the gapped phase preserves certain symmetries, then so too do the resulting boundary conditions.

\para
In some ways, this set-up is more striking. The $\mathbb{Z}_2 \times \mathbb{Z}_2$ symmetry of the gapped phase of 8 Majorana fermions means that both left- and right-moving fermion numbers are preserved mod 2. This is in stark contrast to the familiar boundary conditions which reflect a left-moving fermion into a right-moving fermion.

\para
There are also reasons to think that the boundary approach may be more fruitful in higher dimensions. Much of the work on symmetric mass generation, including \cite{ep,you1,youtoo,simon,ayyar,slagle,juven,juven2}, attempts to gap theories defined on a lattice. In higher dimensions, this is typically achieved by turning on an irrelevant operator and cranking it up to the lattice scale. This is not an option in the continuum, and one must be more creative when trying to understand symmetric mass generation in the language of quantum field theory. (See \cite{you1,youtoo} for one interesting possibility.) In contrast, it may be more straightforward to understand the symmetry-preserving boundary conditions in higher dimensions.

\subsubsection*{The Symmetries of the Boundary}

More interesting boundary conditions typically require the introduction of some boundary degrees of freedom. However, at suitably low energies the resulting physics is elegantly captured in the framework of boundary conformal field theory. In particular, in $d=1+1$ dimensions, the boundary condition can be described by a boundary state, first introduced by Cardy \cite{john}, with the remnant of the boundary degrees of freedom encoded in the boundary central charge \cite{afflud}. The application of boundary conformal field theory to SPT phases was initiated in \cite{shin1,shin2}.

\para
In this paper, we study a large class of conformal boundary states that can be imposed on $2N$ Majorana fermions in $d=1+1$ dimensions. In the absence of a boundary, the Majorana fermions enjoy an $SO(2N)_L \times SO(2N)_R$ chiral symmetry\footnote{The symmetry is, in fact $O(N)_L \times O(N)_R$. The extra elements flip the sign of a single Majorana fermion, and should not be confused with the $\mathbb{Z}_2 \times \mathbb{Z}_2$ chiral fermion parity that is our primary interest.}, which includes the $\mathbb{Z}_2 \times \mathbb{Z}_2$ chiral fermion parity symmetry as its center. We restrict ourselves to boundary states that preserve an anomaly-free $U(1)^N$ subgroup of the maximal torus of $SO(2N)_L \times SO(2N)_R$.

\para
Specifically, we pair the $2N$ Majorana fermions into $N$ Dirac fermions. We assign integer-valued charges $Q_{i \alpha}$ to the left-moving Weyl fermions, and charges $\bar{Q}_{i \alpha}$ to the right-moving Weyl fermions, where $i = 1, \dots, N$ labels the fermion, and $\alpha = 1, \dots, N$ labels the $U(1)$ symmetry. This choice of $U(1)^N$ symmetry is non-anomalous provided
\be \sum_{i=1}^N Q_{\alpha i} Q_{\beta i} = \sum_{i=1}^N \bar{Q}_{\alpha i} \bar{Q}_{\beta i} \nn\ee
It is straightforward to construct boundary states preserving such a chiral symmetry. Early examples include \cite{sagi,juan}. A number of properties of the general class of states have been explored in \cite{us,us2}.

\para
The main question that we would like to address in this paper is: for what choices of $Q$ and $\bar{Q}$ is the chiral $\mathbb{Z}_2\times \mathbb{Z}_2$ fermion parity symmetry restored?

\para
Along the way, we found it useful to also address a related question: for what choices of $Q$ and $\bar{Q}$ is there an enhancement of the preserved $U(1)^N$ to a non-abelian subgroup of $SO(2N)_L\times SO(2N)_R$? Combining these two questions, the purpose of this paper is to describe the full symmetry preserved by a boundary state that is characterised by the chiral charges $Q$ and $\bar{Q}$.

\subsection*{A Statement of our Results}

The answers to our questions above are straightforward to state, and somewhat fiddly to prove. The enhanced symmetries are determined by the rational, orthogonal matrix,
\be \mathcal{R}_{ij} = (\bar{Q}^{-1})_{i\alpha} Q_{\alpha j} \nn\ee
and the associated charge lattice, defined by
\be \LambdaR \coloneqq \mathbb{Z}^N \cap \mathcal{R}^{-1} \mathbb{Z}^N \label{lambdar}\ee
Lattices of this kind, which are the intersection of a lattice with a rotated version of itself, are sometimes referred to as \emph{coincidence site lattices} \cite{baake}. In the present context, the lattice $\LambdaR$ captures the difference between the charges carried by the left- and right-moving fermions. For example, when the charges are equal, with $Q = \bar{Q}$ so $\mathcal{R} = \mathds{1}$, the lattice is simply $\LambdaR = \mathbb{Z}^N$. For boundary states in which the left- and right-moving charges differ, the associated lattice $\LambdaR$ becomes sparser.

\para
Our primary goal is to determine the choices of $Q$ and $\bar{Q}$ that preserve $\mathbb{Z}_2 \times \mathbb{Z}_2$ chiral fermion number. We will show that the boundary condition has such a property if and only if $\LambdaR$ is an even lattice, i.e. the length-squared of any lattice vector is an even integer.

\para
We further show that coincidence site lattices \eqn{lambdar} can be even only when $N$ is divisible by 4. In this way, we reproduce the $\mathbb{Z}_8$ classification of interacting SPT phases in $d=2+1$ dimensions \cite{fk,ryugso,qi,ryu2}, albeit from a rather unconventional perspective.

\para
The simplest boundary state that preserves $\mathbb{Z}_2 \times \mathbb{Z}_2$ chiral symmetry has $2N=8$ Majorana fermions and was constructed long ago by Maldacena and Ludwig to describe the scattering of fermions off a monopole. The relationship of this state to SPT phases was recognised in \cite{shin2}. Our work extends this to all such boundary conditions preserving a $U(1)^N$ symmetry.

\para
As advertised above, along the way we will also need to understand the enhanced non-abelian symmetry. We will show that any such symmetry can be observed by its root system $\DeltaR$ nestled inside $\LambdaR$ such that
\be \DeltaR \coloneqq \big\{ \, \lambda \in \LambdaR \, : \, |\lambda|^2 = 2 \, \big\} \label{deltar}\ee
The simplest, trivial example occurs for a non-chiral boundary condition, which has $Q = \bar{Q}$ and so $\mathcal{R} = \mathds{1}$. In this case, the enhanced symmetry is $SO(2N)_V \subset SO(2N)_L \times SO(2N)_R$. There are, however, a number of less trivial examples. For example the Maldacena-Ludwig boundary state, which preserves $\mathbb{Z}_2 \times \mathbb{Z}_2$, has $\mathcal{R} \neq \mathds{1}$ but also preserves an $SO(8)$. (Indeed, the state was originally constructed to have this property using triality of $SO(8)$.)

\subsubsection*{Plan of the Paper}

We start in Section \ref{contsec} with a brief review of the construction of boundary states preserving chiral symmetries. We then proceed to derive the criterion \eqn{deltar} for the emergence of non-abelian symmetries preserved by the boundary state.

\para
In Section \ref{discretesec}, we turn to the question of $\mathbb{Z}_2 \times \mathbb{Z}_2$ chiral fermion parity. We will first show that this is only a symmetry if it is already a subgroup of the original $U(1)^N$ preserved by the boundary state. Using this result, we then deduce that $N$ must be divisible by 4, and construct all sets of charges $Q$, $\bar{Q}$ with this property.

\section{Continuous Symmetries} \label{contsec}

In this section, we introduce the boundary states that preserve a general, chiral $U(1)^N$ symmetry, before deriving the criterion \eqn{deltar} for the emergence of a larger non-abelian symmetry.

\subsection{The Boundary State} \label{reviewsec}

\begin{figure}[t]
\centering
\includegraphics[scale=1.2]{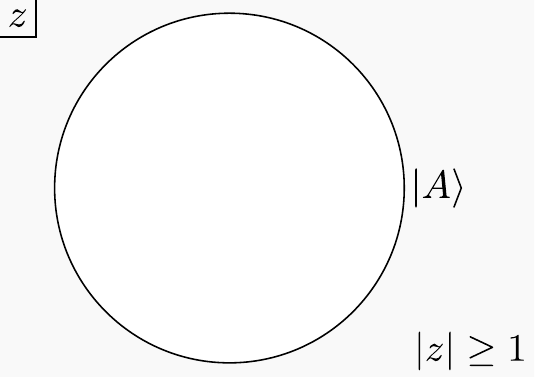}
\caption{The bulk region $|z| \geq 1$, and boundary state $\ket{A}$ at $|z| = 1$.}
\label{theonlyfigure}
\end{figure}

We work with $2N$ Majorana fermions. The set-up described in the introduction, with left- and right-moving fermions moving on a line, can be mapped to a problem in Euclidean space, with action
\be S = \frac{1}{4\pi} \int \! dz d\bar{z} \left( \chi_i \bar{\partial} \chi_i + \bar{\chi}_i \partial \bar{\chi}_i \right) \nn\ee
where we work with complex coordinate $z = x + i \tau$. After a conformal transformation, we can map the boundary from the $\tau$-axis, to the circle $|z| = 1$. The fermions live on the domain $|z| \geq 1$, as shown in Figure \ref{theonlyfigure}.

\para
The advantage of performing the conformal transformation is that, if we now perform radial quantisation, the boundary condition at $|z|=1$ is encoded in a boundary state $\ket{A}$. In the context of SPT phases, these boundary states were discussed in \cite{shin1,shin2}, and a number of their properties explored in \cite{us,us2}. Here we provide only the minimal details necessary to tell our story.

\para
The boundary states depend on the preserved $U(1)^N$ charges $Q$ and $\bar{Q}$, together with $N$ phases $\theta_i$. The charges arise in the guise of the rational, orthogonal matrix $\mathcal{R} = \bar{Q}^{-1} Q$ and the associated lattice \eqn{lambdar}. The most general boundary state takes the form
\be \ket{\theta; \mathcal{R}} \coloneqq g_\mathcal{R} \sum_{\lambda \in \LambdaR} e^{i \gamma_\mathcal{R}(\lambda)} e^{i \theta \cdot \lambda} \, \exp \! \left( -\sum_{n=1}^\infty \frac{1}{n} \mathcal{R}_{ij} \, \bar{J}_{i, -n} J_{j,-n} \right) \ket{\lambda, -\mathcal{R} \lambda} \label{bstate}\ee
This expression involves a number of new ingredients, which we now explain.
\begin{itemize}

\item The currents $J_{i,n}$ and $\bar{J}_{i,n}$ are associated to the $U(1)$ fermion number symmetries for the $N$ left- and $N$ right-moving Weyl fermions\footnote{The $J_{i,n}$ are defined as the modes of $J_i(z) \coloneqq i \chi_{2i-1}(z) \chi_{2i}(z)$.}. Here $i = 1 \dots N$ labels the Weyl fermion, while $n \in \mathbb{Z}$ labels the mode. The currents obey the oscillator algebra $[J_{i,n}, J_{j,m}] = n \delta_{ij} \delta_{n+m,0}$. The negative modes $J_{i,-n}$ act as raising operators and the positive modes $J_{i,n}$ act as lowering operators, where $n \geq 1$.

\item The states $\ket{\lambda, \bar{\lambda}}$ comprise an infinite set of highest weight states for the $U(1)^N_L \times U(1)^N_R$ current algebra, labelled by $\lambda, \bar{\lambda} \in \mathbb{Z}^N$. They are eigenstates of the current zero modes, with eigenvalues $J_{i,0} = \lambda_i$ and $\bar{J}_{i,0} = \bar{\lambda}_i$. Not all highest weight states contribute to the boundary state: only those with $\ket{\lambda, \bar{\lambda} = -\mathcal{R} \lambda}$ appear, where both $\lambda$ and $\bar{\lambda}$ are integer vectors by virtue of the fact that $\lambda \in \LambdaR$.

There is, it turns out, a phase ambiguity in defining these states. This ambiguity was unimportant in previous applications of this state, but will prove important here. We will fix it in Section \ref{contsecproof}.

The negative modes of the currents act on the highest weight state in \eqn{bstate} to create a coherent state, known as an \emph{Ishibashi state}.

\item The boundary state depends on a set of $N$ phases $\theta_i$. These phases appear only in the exponential $e^{i \theta \cdot \lambda}$ where $\lambda$ takes values in $\LambdaR$, which means that $\theta$ naturally takes values in the torus $\mathbb{R}^N / 2 \pi \LambdaR^\star$. This is the moduli space of boundary states for a given $\mathcal{R}$.

\item Each state $\lambda \in \LambdaR$ in the sum is weighted by a phase $e^{i \gamma_\mathcal{R}(\lambda)}$. This, it turns out, is necessary to ensure locality but is otherwise an annoying technicality. For the most part, it plays no role but will make a brief appearance in a calculation in Section~\ref{contsecproof}.

\item The overall normalisation of the boundary state is the Affleck-Ludwig central charge, $g_\mathcal{R} = \langle 0, 0 | \theta; \mathcal{R} \rangle$ \cite{afflud}. It is given in terms of the volume of the primitive unit cell of the lattice: $g_\mathcal{R} = \sqrt{\vol(\LambdaR)}$ \cite{bachas,us}.

\end{itemize}

\subsection{Enhanced Continuous Symmetries} \label{contsecproof}

In this section, we answer the second of the two questions posed in the introduction: what is the full sub-algebra of $\mathfrak{so}(2N)_L \times \mathfrak{so}(2N)_R$ left unbroken by the boundary state $\ket{\theta; \mathcal{R}}$?

\para
To say that an infinitesimal symmetry is unbroken at a boundary is to say that its corresponding Noether current has vanishing flux into the boundary. For us, the symmetry under consideration is $(A, \bar{A}) \in \mathfrak{so}(2N)_L \times \mathfrak{so}(2N)_R$, and the condition reads
\be \left( z J_A(z) + \bar{z} \bar{J}_{\bar{A}}(\bar{z}) \right) \ket{\theta; \mathcal{R}} = 0 \qquad \text{for all } |z| = 1 \label{noflowcurrents}\ee
Here the $J_A(z)$ is the $\mathfrak{so}(2N)_L$ current associated to the generator $A$, and $\bar{J}_{\bar{A}}(\bar{z})$ is the $\mathfrak{so}(2N)_R$ current associated to the generator $\bar{A}$.\footnote{The currents are defined by $J_A(z) \coloneqq \frac{1}{2} A_{ij} \chi_i(z) \chi_j(z)$, where the generator $A$ is regarded as a real, $2N \times 2N$ antisymmetric matrix $A_{ij}$.} Note that these non-abelian currents are distinguished from their Cartan counterparts $J_i$ and $\bar{J}_i$ only by their index.

\para
Our goal is to find all solutions $(A, \bar{A})$ to the above equation. However, doing this directly by computing the action of the currents on the boundary state turns out to be a bit of a pain. Instead, we will take a slightly different approach. We will first show, using algebraic arguments and anomalies, that the problem of checking \eqn{noflowcurrents} for general $(A, \bar{A})$ can be reduced to checking it for a certain finite set of special generators. Then we carry out the check for these special generators directly.

\para
There is one other simplification that is important to make before we start. Instead of looking for solutions with $A, \bar{A}$ in $\mathfrak{so}(2N)$, we will take them to lie in $\mathfrak{so}(2N)_\mathbb{C}$. Solving the complexified problem turns out to be technically easier, and besides, we can always recover the answer to the original question by imposing a reality condition on $A, \bar{A}$.

\subsubsection*{Anomalies}

Our first job is to show that the set of solutions $(A, \bar{A})$ to \eqn{noflowcurrents} forms an anomaly-free subalgebra. We begin by recasting \eqn{noflowcurrents} as an equivalent algebraic condition, written in terms of modes,
\be \left( J_{A, n} + \bar{J}_{\bar{A}, -n} \right) \ket{\theta; \mathcal{R}} = 0 \qquad \text{for all } n \in \mathbb{Z} \label{noflowmodes}\ee
Suppose that we have two solutions $(A, \bar{A})$ and $(B, \bar{B})$. Clearly, we have
\be [ J_{A, n} + \bar{J}_{\bar{A}, -n} , J_{B, m} + \bar{J}_{\bar{B}, -m} ] \, \ket{\theta; \mathcal{R}} = 0 \qquad \text{for all } n,m \in \mathbb{Z} \nn\ee
Using the fact that the modes obey the $\widehat{\mathfrak{so}}(2N)_1$ algebra
\be [J_{A,n}, J_{B,m}] = J_{[A,B], n+m} + n \delta_{n+m,0} \, K(A,B) \nn\ee
with $K(A, B) = \frac{1}{2} \tr(AB)$ the correctly-normalised Killing form, we can simplify the previous equation to
\be \Big( \underline{J_{[A,B], n+m} + \bar{J}_{[\bar{A},\bar{B}], -(n+m)}} + n \delta_{n+m,0} \underline{(K(A, B) - K(\bar{A}, \bar{B}))} \Big) \, \ket{\theta; \mathcal{R}} = 0 \nn\ee
Since this must hold for all $n,m$, both underlined terms must vanish separately. The first shows that the solutions close into an algebra, while the second forces the vanishing of the mixed 't Hooft anomaly between any two solutions
\be K(A, B) - K(\bar{A}, \bar{B}) = 0 \nn\ee
which is what we wanted to show.

\para
Next, we show that the set of solutions to \eqn{noflowcurrents}, or the \emph{unbroken subalgebra}, has a second important property: whenever $(A, \bar{A})$ is a solution, so is its complex conjugate $(A^*, \bar{A}^*)$. To see this, we cook up a judicious choice of antilinear operation $\mathcal{T}$ defined by the stipulations
\be
\mathcal{T} \ket{\lambda, \bar{\lambda}} &= (-1)^{\lambda^2} \ket{-\lambda, -\bar{\lambda}} \nn\\
\mathcal{T} J_{i, n} \mathcal{T}^{-1} &= -J_{i, n} \nn\\
\mathcal{T} \bar{J}_{i, n} \mathcal{T}^{-1} &= -\bar{J}_{i, n} \nn
\ee
Under this operation, the boundary states are invariant,
\be \mathcal{T} \ket{\theta; \mathcal{R}} = \ket{\theta; \mathcal{R}} \nn\ee
while the non-abelian currents transform as
\be
\mathcal{T} J_{A}(z) \mathcal{T}^{-1} = J_{A^*}(z^*)
\ \ \ \text{and} \ \ \
\mathcal{T} \bar{J}_{A}(\bar{z}) \mathcal{T}^{-1} = \bar{J}_{A^*}(\bar{z}^*)
\nn\ee
Acting on \eqn{noflowcurrents} with $\mathcal{T}$, this then establishes the claim\footnote{To give more details, the first transformation uses \eqn{bstate} and the identities $e^{-i \gamma_{\mathcal{R}}(-\lambda)} = e^{-i \gamma_{\mathcal{R}}(\lambda)} = (-1)^{\lambda^2} e^{i \gamma_{\mathcal{R}}(\lambda)}$ which follow from \eqn{eigamma}. The last two transformations follow from the bosonised form of the currents, both diagonal \eqn{diagonalcurrents} and off-diagonal \eqn{bosonisedcurrents}.}.

\para
The previous two results place strong constraints on the structure of the unbroken subalgebra, and ultimately force it to take the form
\be \{ \, (A, \, \bar{A} = \phi(A)) \, : \, A \in \mathfrak{g} \, \} \label{constrainedform}\ee
Here $\mathfrak{g}$ is some subalgebra of $\mathfrak{so}(2N)_\mathbb{C}$, describing the allowed holomorphic parts $A$ of the symmetries, while $\phi$ is a map $\mathfrak{g} \rightarrow \mathfrak{so}(2N)_\mathbb{C}$ which sends them to their corresponding antiholomorphic parts $\bar{A}$. The map $\phi$ must be both a homomorphism of Lie algebras, and an isometry with respect to the Killing form $K$, in order that \eqn{constrainedform} describe an anomaly-free subalgebra.

\para
To demonstrate the assertion of the last paragraph, note that the unbroken subalgebra cannot contain any purely chiral elements $(A, 0)$. For if it did, then it would also contain $(A^*, 0)$, and then the anomaly condition $K(A, A^*) = 0$ would force $A = 0$, since the Killing form is negative definite. Therefore, for every generator $A \in \mathfrak{so}(2N)_\mathbb{C}$, there can be at most a single $\bar{A} \in \mathfrak{so}(2N)_\mathbb{C}$ such that $(A, \bar{A})$ is in the unbroken subalgebra. This forces the algebra to take the form \eqn{constrainedform}, as claimed.

\subsubsection*{The Existing Abelian Symmetry}

Until now, we haven't actually used any properties of the boundary state $\ket{\theta; \mathcal{R}}$ itself. Here we take into account the property that it preserves a $U(1)^N$ symmetry. This is encoded by the identity
\be (\mathcal{R}_{ij} J_{j, n} + \bar{J}_{i, -n}) \, \ket{\theta; \mathcal{R}} = 0 \nn\ee
as can easily be verified starting from the definition \eqn{bstate}. Our goal is to determine the resulting constraints that this imposes on the algebra \eqn{constrainedform}. To do this, we use the relation
\be J_i(z) = J_{H_i}(z) \label{diagonalcurrents}\ee
between the $\mathfrak{u}(1)^N$ and $\mathfrak{so}(2N)$ currents, where the generators $H_i$ are the following choice of basis for the Cartan subalgebra,
\be
H_1 = \left(\begin{array}{ccc} 0 & i \\ -i & 0 \\ & & \ddots \end{array}\right)
\qquad \dots \qquad
H_N = \left(\begin{array}{ccc} \ddots \\ & 0 & i \\ & -i & 0 \end{array}\right)
\nn\ee
The existence of the $U(1)^N$ symmetry can then be recast as
\be (\mathcal{R}_{ij} J_{H_j, n} + \bar{J}_{H_i, -n}) \, \ket{\theta; \mathcal{R}} = 0 \nn\ee
This equation now takes the same form as \eqn{noflowmodes}, and states that the symmetry $(A, \bar{A}) = (\mathcal{R}_{ij} H_j, H_i)$ is preserved for each $i$, or equivalently, that $(H_i, \mathcal{R}_{ji} H_j)$ is. We can now read off the constraints on the algebra \eqn{constrainedform} that we were looking for:
\begin{enumerate}
\item $\mathfrak{g}$ contains the Cartan subalgebra: $H_i \in \mathfrak{g}$ for all $i$.
\item $\phi$ is uniquely determined on the Cartan subalgebra: $\phi(H_i) = \mathcal{R}_{ji} H_j$.
\end{enumerate}

\subsubsection*{Algebraic Constraints}

The above two constraints allow us to reduce the task of solving \eqn{noflowcurrents} for $(A, \bar{A})$ to checking a finite list of candidates. The candidates are
\be (A, \bar{A}) = (E_\alpha, e^{i \chi} E_{R \alpha}) \nn\ee
where $\alpha$ ranges over the finite set $\DeltaR$ introduced in \eqn{deltar}, and $\chi$ is a phase to be determined as part of the test. The enhanced symmetry is then given by the linear span of all the generators of the above kind that pass the test\footnote{This statement makes free use of the identification between the roots of $\mathfrak{so}(2N)$ and the vectors in $\mathbb{Z}^N$ of length-squared 2. Indeed, given a root $\alpha$, its components $(\alpha_1, \dots, \alpha_N)$ with respect to the Cartan generators $H_i$ are exactly a vector of this form. The subset $\DeltaR \subset \mathbb{Z}^N$ has the property that, for any member $\alpha$, both $\alpha$ and $\mathcal{R} \alpha$ are roots; this ensures the expression for $(A, \bar{A})$ makes sense.}.

\para
To prove this claim, let $A$ be any generator in $\mathfrak{g}$. We can decompose $A$ in the Chevalley basis as
\be A = \sum_{i=1}^N x_i H_i + \sum_{\alpha \in \Delta} x_\alpha E_\alpha \nn\ee
where $\Delta$ is the set of roots of $\mathfrak{so}(2N)$, and the $x_i$ and $x_\alpha$ are sets of complex coefficients. For any $\alpha$ such that $x_\alpha \neq 0$, we can repeatedly take commutators with $H_i$ and form linear combinations to project $A$ onto the single generator $E_\alpha$. Because these operations do not take us outside $\mathfrak{g}$, the result $E_\alpha$ must lie in $\mathfrak{g}$. We learn that $\mathfrak{g}$ is fully determined by asking whether it contains $E_\alpha$ for each $\alpha \in \Delta$, and is given by the span of all such generators.

\para
Next we restrict the possible $\alpha$ that can occur to the set $\DeltaR$. Suppose that $E_\alpha \in \mathfrak{g}$. Using the homomorphism property $[\phi(H_i), \phi(E_{\alpha})] = \phi([H_i, E_\alpha])$ together with the action on the Cartan generators $\phi(H_i) = \mathcal{R}_{ji} H_j$, we deduce that
\be [H_i, \phi(E_\alpha)] = \mathcal{R}_{ij} \alpha_j \, \phi(E_{\alpha}) \nn\ee
This states that $\phi(E_\alpha)$ is proportional to $E_{\mathcal{R} \alpha}$. For this to be possible at all, both $\alpha$ and $\mathcal{R} \alpha$ must be roots. To see what this says about $\alpha$, first note that since both $\alpha$ and $\mathcal{R} \alpha$ are integer vectors, $\alpha$ must lie in $\LambdaR$. The additional requirement they be roots is that $|\alpha|^2 = 2$, which further restricts $\alpha$ to lie in $\DeltaR$, as we wanted to show.

\para
Finally, given $\alpha \in \DeltaR$, the proportionality constant in $\phi(E_\alpha) \propto E_{\mathcal{R} \alpha}$ must actually be a phase. This follows because $\phi$ is an isometry with respect to the Killing form. Denoting this phase by $e^{i \chi}$, we recover the result claimed at the start.

\subsubsection*{Explicitly Checking the Generators}

The only remaining task is to test the generators identified at the start of the last section to see whether they satisfy \eqn{noflowcurrents}. We will find that in fact they all do. That is, there are no further obstructions to the existence of enhanced symmetries beyond those we have already identified.

\para
This section is necessarily slightly more technical than the rest. Before proceeding, we'll finally need to face up to a number of points whose discussion we've been trying to outrun for some time:
\begin{itemize}

\item The phase ambiguity of the $\ket{\lambda, \bar{\lambda}}$.

In equation \eqn{bstate} we introduced the states $\ket{\lambda, \bar{\lambda}}$, but did not specify how their phases were to be fixed. This is done by bosonisation. First we pair the $2N$ real fermions $\chi_i$ up into $N$ complex fermions $\psi_i(z) = \frac{1}{\sqrt{2}}(\chi_{2i-1}(z) + i \chi_{2i}(z))$. These are then expressed in terms of the abelian currents as
\be \psi_i(z) &= F_i \, t_i \, z^{-\lambda_i} \exp \! \left( -\sum_{n=1}^\infty \frac{z^n}{n} J_{i, -n} \right) \exp \! \left( \sum_{n=1}^\infty \frac{z^{-n}}{n} J_{i, n} \right) \nn\ee
Here we have introduced two new ingredients, which will pervade what follows: a set of ladder operators $F_i$, and cocycles $t_i$. The former are defined to be the operators that move between the highest-weight states as
\be F_i \ket{\lambda, \bar{\lambda}} = \ket{\lambda - e_i, \bar{\lambda}} \nn\ee
while the latter multiply them by a phase
\be t_i \ket{\lambda, \bar{\lambda}} = t_i(\lambda, \bar{\lambda}) \, \ket{\lambda, \bar{\lambda}} \nn\ee
All this of course comes with an identical antiholomorphic twin. The phase ambiguity of the $\ket{\lambda, \bar{\lambda}}$ now amounts to the freedom to choose different conventions for the cocycles. We fix it by making the choice
\be t_i(\lambda, \bar{\lambda}) = (-1)^{\sum_{j=1}^{i-1} \lambda_j} \qquad \bar{t}_i(\lambda, \bar{\lambda}) = (-1)^{\sum_{j=1}^{i-1} \bar{\lambda}_j + \sum_{j=1}^N \lambda_j} \label{cocycles}\ee
Note that this is one of the few places where there is \emph{not} a perfect symmetry between holomorphic and antiholomorphic sectors.

\item The phases $e^{i \gamma_\mathcal{R}(\lambda)}$.

A further set of phases $e^{i \gamma_\mathcal{R}(\lambda)}$ must be included in \eqn{bstate} on grounds of locality. A detailed explanation can be found in \cite{us}; here we simply recount their definition. First introduce a mod-2 valued bilinear form $f_{\mathcal{R}}(q, p)$ on $\LambdaR$, defined by
\be f_{\mathcal{R}}(q, p) \coloneqq \sum_{i=1}^N p_i \sum_{j=1}^{i-1} q_j + \sum_{i=1}^N (\mathcal{R} p)_i \bigg(\sum_{j=1}^{i-1} (\mathcal{R} q)_j + \sum_{j=1}^N q_j \bigg) \mod 2 \label{fdefn}\ee
This is a symmetric bilinear form, and enjoys the property $f_{\mathcal{R}}(\lambda, \lambda) = \lambda^2 \text{ mod } 2$. Next pick an arbitrary lift $\hat{f}_{\mathcal{R}}(q, p)$ to a mod-4 valued symmetric bilinear form. Then the phases are given by
\be e^{i \gamma_\mathcal{R}(\lambda)} \coloneqq i^{\hat{f}_{\mathcal{R}}(\lambda, \lambda)} \label{eigamma}\ee
The ambiguity in the choice of lift results in the freedom to shift
\be e^{i \gamma_\mathcal{R}(\lambda)} \rightarrow e^{i \gamma_\mathcal{R}(\lambda)} (-1)^{s \cdot \lambda} \label{signambiguity}\ee
by an arbitrary element $s \in \LambdaR^\star$. This ambiguity cannot be fixed in any canonical way, and we simply have to live with it.

\item The $\mathfrak{so}(2N)$ roots, generators and currents.

We'll also need to describe the root system $\Delta$ of $\mathfrak{so}(2N)$ a little more explicitly. Each root of $\mathfrak{so}(2N)$ is labelled by a pair of integers $1 \leq i < j \leq N$ and a pair of signs $s, t = \pm 1$, and is represented by the $N$-component vector
\be \alpha_{i,j,s,t} = (0, \dots, 0, \stackrel{\begin{smallmatrix} i \\ \downarrow \end{smallmatrix}}{s}, 0, \dots, 0, \stackrel{\begin{smallmatrix} j \\ \downarrow \end{smallmatrix}}{t}, 0, \dots, 0) \nn\ee
The corresponding generator is the $2N \times 2N$ matrix
\be
E_{i,j,s,t} \; = \quad \frac{1}{2} \quad
\begin{blockarray}{cccccccc}
 \hdots & 2i-1 & 2i & \hdots & 2j-1 & 2j & \hdots & \\
 \begin{block}{(ccccccc)l}
   & & & & & & & \vdots \\
   & & & & 1 & it & & 2i-1 \\
   & & & & is & -st & & 2i \\
   & & & & & & & \vdots \\
   & -1 & -is & & & & & 2j-1 \\
   & -it & st & & & & & 2j \\
   & & & & & & & \vdots \\
 \end{block}
\end{blockarray}
\nn\ee
and the corresponding bosonised currents are
\be J_{i,j,s,t}(z) = (F_i)^s (F_j)^t \, t_i t_j \, z^{-\alpha_k \lambda_k} \, \mathscale{0.85}{\exp \! \left( -\alpha_k \sum_{n=1}^\infty \frac{z^n}{n} J_{k, -n} \right) \exp \! \left( \alpha_k \sum_{n=1}^\infty \frac{z^{-n}}{n} J_{k, n} \right)} \label{bosonisedcurrents}\ee
with a similar expression for the antiholomorphic currents $\bar{J}_{i,j,s,t}(\bar{z})$.

\end{itemize}

\vspace{0.2em}

\para
We now proceed to carry out the check. Let $\alpha$ be any root in $\DeltaR$. Since $\alpha$ and $\mathcal{R} \alpha$ are both roots, they can be written in the form
\be \alpha = \alpha_{i,j,s,t} \qquad \mathcal{R} \alpha = \alpha_{i',j',s',t'} \nn\ee
for some choices of the various indices. The condition \eqn{noflowcurrents} for our candidate generator $(E_\alpha, e^{i \chi} E_{R \alpha})$ to be preserved then takes the form
\be \big( J_{i,j,s,t}(z) + e^{i \chi} z^{-2} \bar{J}_{i',j',s',t'}(\bar{z}) \big) \ket{\theta; \mathcal{R}} \stackrel{?}{=} 0 \nn\ee
where $|z| = 1$. Next, we plug in the definition of the boundary state \eqn{bstate} and bosonised currents \eqn{bosonisedcurrents}. The condition becomes
\be
\sum_{\lambda \in \LambdaR} &(t_i t_j|_{\lambda,-\mathcal{R} \lambda}) e^{i \gamma_\mathcal{R}(\lambda)} e^{i \theta \cdot \lambda} z^{-\alpha \cdot \lambda} \, \mathscale{0.85}{\exp \! \left( -\alpha_k \sum_{n=1}^\infty \frac{z^n}{n} J_{k, -n} \right)} \nn\\
&\quad \mathscale{0.85}{\exp \! \left( \alpha_k \sum_{n=1}^\infty \frac{z^{-n}}{n} J_{k, n} \right) \exp \! \left( -\sum_{n=1}^\infty \frac{1}{n} \mathcal{R}_{ij} \, \bar{J}_{i, -n} J_{j,-n} \right)} \ket{\lambda-\alpha, -\mathcal{R} \lambda} \nn\\[1em]
{} \; + \; e^{i \chi} \sum_{\lambda \in \LambdaR} &(\bar{t}_{i'} \bar{t}_{j'} |_{\lambda,-\mathcal{R} \lambda}) e^{i \gamma_\mathcal{R}(\lambda)} e^{i \theta \cdot \lambda} z^{-\alpha \cdot \lambda - 2} \, \mathscale{0.85}{\exp \! \left( -(\mathcal{R} \alpha)_k \sum_{n=1}^\infty \frac{z^{-n}}{n} \bar{J}_{k, -n} \right)} \nn\\
&\quad \mathscale{0.85}{\exp \! \left( (\mathcal{R} \alpha)_k \sum_{n=1}^\infty \frac{z^n}{n} \bar{J}_{k, n} \right) \exp \! \left( -\sum_{n=1}^\infty \frac{1}{n} \mathcal{R}_{ij} \, \bar{J}_{i, -n} J_{j,-n} \right)} \ket{\lambda, -\mathcal{R} (\lambda + \alpha)}
\; \stackrel{?}{=} \; 0
\nn\ee
A short calculation shows that the oscillator parts of the above two sums are equal. This means that the exponentials of oscillators can be dropped. We can also shift the argument of the second sum by $\lambda \rightarrow \lambda - \alpha$ to make the sums look more alike. This is allowed because $\alpha \in \LambdaR$. After the dust has settled, we are left with
\be (t_i t_j|_{\lambda, -\mathcal{R} \lambda}) \, e^{i \gamma_\mathcal{R}(\lambda)} + e^{i (\chi - \theta \cdot \alpha)} \, (\bar{t}_{i'} \bar{t}_{j'} |_{\lambda - \alpha, -\mathcal{R} (\lambda - \alpha)}) \, e^{i \gamma_\mathcal{R}(\lambda - \alpha)} \; \stackrel{?}{=} \; 0 \nn\ee
This equation is required to hold for each $\lambda \in \LambdaR$.

\para
To verify that it actually does hold, we appeal to a bunch of identities satisfied by the various objects that arise:
\begin{itemize}

\item From definition \eqn{cocycles}, and the fact $\mathcal{R} \alpha = \alpha_{i',j',s',t'}$,
\be (\bar{t}_{i'} \bar{t}_{j'} |_{\lambda - \alpha, -\mathcal{R} (\lambda - \alpha)}) = -(\bar{t}_{i'} \bar{t}_{j'} |_{\lambda, -\mathcal{R} \lambda}) \nn\ee

\item From definition \eqn{eigamma}, bilinearity of $\hat{f}$, and the fact $\hat{f}_\mathcal{R} = f_\mathcal{R}$ mod 2,
\be e^{i \gamma_\mathcal{R}(\lambda - \alpha)} = e^{i \gamma_\mathcal{R}(\lambda)} e^{i \gamma_\mathcal{R}(\alpha)} (-1)^{f_\mathcal{R}(\lambda, \alpha)} \nn\ee

\item From definitions \eqn{cocycles} and \eqn{fdefn},
\be (-1)^{f_\mathcal{R}(\lambda, \alpha)} = (t_i t_j |_{\lambda, -\mathcal{R} \lambda}) (\bar{t}_{i'} \bar{t}_{j'} |_{\lambda, -\mathcal{R} \lambda}) \nn\ee

\end{itemize}
Combining the above three identities shows that the condition is satisfied, with the phase $\chi$ given by $e^{i \chi} = e^{i \theta \cdot \alpha} e^{-i \gamma_\mathcal{R}(\alpha)}$. The factor $e^{i \gamma_\mathcal{R}(\alpha)}$ can be set to one using the ambiguity \eqn{signambiguity}. To see this, first note that $(e^{i \gamma_\mathcal{R}(\alpha)})^2 = (-1)^{|\alpha|^2 = 2} = 1$, so $e^{i \gamma_\mathcal{R}(\alpha)}$ is a sign. It can be argued that these signs fit together in a consistent way, in the sense that $e^{i \gamma_\mathcal{R}(\alpha)} = (-1)^{s \cdot \alpha}$ for some $s$. This is precisely of a form that can be absorbed by a shift \eqn{signambiguity}. So without loss of generality, $e^{i \chi} = e^{i \theta \cdot \alpha}$.

\subsubsection*{Summary}

The upshot of this section is that the unbroken subalgebra for $\ket{\theta; \mathcal{R}}$ is spanned by the abelian generators, $(H_i, \mathcal{R}_{ij} H_j)$, with $i=1, \dots, N$, together with the off-diagonal enhanced generators
\be (E_\alpha, e^{i \theta \cdot \alpha} E_{\mathcal{R} \alpha}) \quad : \quad \alpha \in \DeltaR \nn\ee
with $\DeltaR$ consisting of the points $\alpha \in \LambdaR$ with length-squared $\alpha^2 = 2$.

\subsection{Some Examples} \label{contsecproofexamples}

In this section we give a few examples of the criterion. First, we show how it can be used to classify all states with maximal symmetry (i.e. with a preserved subalgebra isomorphic to $\mathfrak{so}(2N)$). Second, we explore a family of states which previously arose in the context of fermions scattering off monopoles and, for that reason, are called \emph{dyon states}. These will go on to play a starring role in Section~\ref{discretesec}.

\para
To begin, suppose that $\ket{\theta; \mathcal{R}}$ has unbroken subalgebra isomorphic to $\mathfrak{so}(2N)$. By the criterion, this happens whenever
\be \mathcal{R} \Delta = \Delta \nn\ee
where $\Delta$ is the root system of $\mathfrak{so}(2N)$. We will classify such matrices $\mathcal{R}$ up to the freedom to shift $\mathcal{R} \rightarrow \mathcal{W}_R \mathcal{R} \mathcal{W}_L$ where $\mathcal{W}_{L}, \mathcal{W}_{R}$ are $N \times N$ signed permutation matrices. (Transforming $\mathcal{R}$ in this way corresponds to acting on the boundary state with an $O(2N)_L \times O(2N)_R$ Weyl group transformation.) Using this freedom, we may assume that $\mathcal{R}$ maps the simple roots to a permutation of themselves. We can then classify $\mathcal{R}$ by considering its induced Dynkin diagram automorphism:
\begin{itemize}

\item
For $N \neq 4$, the Dynkin diagram has a $\mathbb{Z}_2$ automorphism group, whose generator exchanges the two spinor representations. But this automorphism can be undone by acting with a $O(2N)$ Weyl transformation $\mathcal{W}_R$ that is not a $SO(2N)$ Weyl transformation. Therefore, the only possibility in this case is the trivial state
\be \mathcal{R} = \mathds{1} \quad \text{with} \quad g_{\mathcal{R}} = 1 \nn\ee

\item
For $N = 4$, the Dynkin diagram of $\mathfrak{so}(8)$ exhibits a famous triality, and the automorphism group is $\mathbb{S}_3$. In addition to the $\mathbb{Z}_2$ above there are 4 nontrivial triality automorphisms. Under the identifications $\mathcal{R} \sim \mathcal{W}_R \mathcal{R} \mathcal{W}_L$ these collapse to just one,
\be \mathcal{R} = \frac{1}{2} \begin{pmatrix} +1 & -1 & -1 & -1 \\ -1 & +1 & -1 & -1 \\ -1 & -1 & +1 & -1 \\ -1 & -1 & -1 & +1 \end{pmatrix} \quad \text{with} \quad g_\mathcal{R} = \sqrt{2} \label{mlstate}\ee
This state was first constructed by Maldacena and Ludwig \cite{juan} in the context of fermion-monopole scattering.

\end{itemize}

The Maldacena-Ludwig state lies in a sequence of so called \emph{dyon states} that exist for general $N$. They have preserved charges and boundary central charge given by
\be \mathllap{\text{dyon}_N : \hspace{1.5cm}} \mathcal{R}_{ij} = \delta_{ij} - \frac{2}{N} \quad \text{with} \quad g_\mathcal{R} = \sqrt{\tfrac{N}{\gcd(N,2)}} \label{dyons}\ee
For most values of $N$, the enhanced symmetry of the dyon state is $SU(N)_V \times U(1)_A$ and no larger. This symmetry is chiral: the left-moving fermions transform in $\mathbf{N}_{+1}$, while the right-movers transform in $\mathbf{N}_{-1}$.

\para
There are three exceptions to the statement above. When $N=1$ and $N=2$, the dyon states are trivial and preserve the subgroup $SO(2N)_V$. When $N=4$, the dyon state coincides with the Maldacena-Ludwig state, which preserves a $\text{Spin}(8)$ subgroup nontrivially embedded within $SO(8) \times SO(8)$.

\section{Discrete Symmetries} \label{discretesec}

In this section we turn to turn to our primary goal: to determine which charges $Q$ and $\bar{Q}$ admit a $\mathbb{Z}_2 \times \mathbb{Z}_2$ preserving boundary condition.

\para
The $\mathbb{Z}_2 \times \mathbb{Z}_2$ symmetry acts via left and right fermion parity which, in Euclidean space, we continue to denote as $(-1)^{F_L}$ and $(-1)^{F_R}$. Under these symmetries, the ground states transform with charges
\be
(-1)^{F_L} \ket{\lambda, \bar{\lambda}} = (-1)^{\lambda_1 + \dots + \lambda_N} \ket{\lambda, \bar{\lambda}} = (-1)^{\lambda^2} \ket{\lambda, \bar{\lambda}} \nn\\
(-1)^{F_R} \ket{\lambda, \bar{\lambda}} = (-1)^{\bar{\lambda}_1 + \dots + \bar{\lambda}_N} \ket{\lambda, \bar{\lambda}} = (-1)^{\bar{\lambda}^2} \ket{\lambda, \bar{\lambda}} \nn
\ee
while the abelian currents are uncharged. The condition for a boundary state $\ket{\theta; \mathcal{R}}$ to be neutral under both symmetries is simply the requirement that the allowed charge sectors $\ket{\lambda, -\mathcal{R} \lambda}$, with $\lambda \in \LambdaR$, are all neutral. This in turn will hold if the lattice $\LambdaR$ contains only vectors of even length-squared. We can equivalently write this as
\be \LambdaR \subseteq D_N \nn\ee
with $D_N$ the root lattice of $\mathfrak{so}(2N)$. Our goal is to investigate the set of all $\mathcal{R}$ for which this condition holds.

\subsection{Discrete symmetries are continuous}

Our starting point is the following lemma, which applies to the global structure of the symmetry group of the boundary state. We restrict attention to the $U(1)^N \times U(1)^N$ maximal torus of $SO(2N)_L \times SO(2N)_R$, which we parametrise as
\be U(1)^N \times U(1)^N = \big\{ \, (e^{2 \pi i x}, e^{2 \pi i \bar{x}}) \, : \, x, \bar{x} \in \mathbb{R}^N / \mathbb{Z}^N \big\} \nn\ee
Of this, the boundary state $\ket{\theta; \mathcal{R}}$ was designed to preserve the subgroup
\be U(1)_\mathcal{R}^N \coloneqq \big\{ \, ( e^{2 \pi i x}, e^{2 \pi i \mathcal{R} x} ) \, : \, x \in \mathbb{R}^N / \LambdaR \, \big\} \nn\ee
In principle, it could be the case that the boundary state also preserves some extra discrete symmetries that lie in $U(1)^N \times U(1)^N$ but not in $U(1)_\mathcal{R}^N$. We will show that this does not happen.

\paragraph{Claim:}

Any element of $U(1)^N \times U(1)^N$ that leaves $\ket{\theta; \mathcal{R}}$ invariant lies in $U(1)^N_\mathcal{R}$.

\paragraph{Proof:}

We begin with the obvious exact sequence of abelian groups
\[ 0 \longrightarrow \mathbb{R}^N \xrightarrow[\left(\begin{smallmatrix} \mathds{1} \\ \mathcal{R} \end{smallmatrix}\right)]{} \mathbb{R}^N \oplus \mathbb{R}^N \xrightarrow[\left(\begin{smallmatrix} \mathds{1} \, & -\mathcal{R}^{-1} \end{smallmatrix}\right)]{} \mathbb{R}^N \longrightarrow 0 \]
This contains the sub-exact-sequence
\be 0 \longrightarrow \mathbb{Z}^N \cap \mathcal{R}^{-1} \mathbb{Z}^N \longrightarrow \mathbb{Z}^N \oplus \mathbb{Z}^N \longrightarrow \mathbb{Z}^N + \mathcal{R}^{-1} \mathbb{Z}^N \longrightarrow 0 \nn\ee
Taking the quotient of these yields the further exact sequence
\be 0 \longrightarrow \frac{\mathbb{R}^N}{\mathbb{Z}^N \cap \mathcal{R}^{-1} \mathbb{Z}^N} \longrightarrow \frac{\mathbb{R}^N} {\mathbb{Z}^N} \oplus \frac{\mathbb{R}^N}{\mathbb{Z}^N} \longrightarrow \frac{\mathbb{R}^N}{\mathbb{Z}^N + R^{-1} \mathbb{Z}^N} \longrightarrow 0 \nn\ee
In the denominator of the left term, we see the definition $\mathbb{Z}^N \cap \mathcal{R}^{-1} \mathbb{Z}^N = \LambdaR$. Furthermore, taking the dual of this definition yields $\mathbb{Z}^N + \mathcal{R}^{T} \mathbb{Z}^N = \LambdaR^\star$. Because $\mathcal{R}$ is orthogonal, we can replace in this equation $R^T = R^{-1}$. The above exact sequence then takes the final form
\be 0 \longrightarrow \frac{\mathbb{R}^N}{\LambdaR} \longrightarrow \frac{\mathbb{R}^N}{\mathbb{Z}^N} \oplus \frac{\mathbb{R}^N}{\mathbb{Z}^N} \longrightarrow \frac{\mathbb{R}^N}{\LambdaR^\star} \longrightarrow 0 \label{exact}\ee
This exact sequence also appeared in \cite{juven}. The various groups appearing in this sequence all have simple interpretations:
\begin{itemize}

\item The middle group is the group of all $U(1)^N \times U(1)^N$ transformations.

\item The left group is the preserved $U(1)^N_\mathcal{R}$ subgroup, with the map into $U(1)^N \times U(1)^N$ simply corresponding to the inclusion map.

\item The right group is the obstruction of a $U(1)^N \times U(1)^N$ transformation to be a symmetry of $\ket{\theta; \mathcal{R}}$. To see this, we recall the fact that such a transformation acts on the ground states as
\be (e^{2 \pi i x}, e^{2 \pi i \bar{x}}) \; : \; \ket{\lambda, \bar{\lambda}} \; \rightarrow \; e^{2 \pi i (x \cdot \lambda + \bar{x} \cdot \bar{\lambda})} \ket{\lambda, \bar{\lambda}} \nn\ee
In order for $\ket{\theta; \mathcal{R}}$ to be invariant, all the ground states that occur in it must be neutral. These ground states are $\ket{\lambda, -\mathcal{R} \lambda}$ for $\lambda \in \LambdaR$. Therefore, we require
\be (x - \mathcal{R}^T \bar{x}) \cdot \lambda \in \mathbb{Z} \qquad \text{for all } \lambda \in \LambdaR \nn\ee
This failure of the quantity $x - \mathcal{R}^T \bar{x}$ to lie in $\LambdaR^\star$ therefore measures the obstruction for the transformation to be a symmetry, as claimed.

\end{itemize}
Exactness of \eqn{exact} at the middle group then gives the desired conclusion, that a $U(1)^N \times U(1)^N$ transformation is a symmetry if and only if it lives in $U(1)^N_\mathcal{R}$.

\subsection{The $\mathbb{Z}_2 \times \mathbb{Z}_2$ symmetry} \label{sec:z2z2critera}

From now on, we specialise to the $\mathbb{Z}_2 \times \mathbb{Z}_2$ symmetry. As we already have seen, the condition for this to be preserved is
\be \lambda^2 = 0 \text{ mod } 2 \qquad \text{for all } \lambda \in \LambdaR \nn\ee
Applying the result of the previous section, we see that this is equivalent to either of the following two stronger statements:
\be \exists\ \lambda \in \LambdaR &\qquad \text{s.t.} \qquad \begin{aligned} \lambda_i &= 1 \text{ mod } 2 \\ (\mathcal{R} \lambda)_i &= 0 \text{ mod } 2 \end{aligned} \nn\ee
and
\be \exists \ \lambda \in \LambdaR &\qquad \text{s.t.} \qquad \begin{aligned} \lambda_i &= 0 \text{ mod } 2 \\ (\mathcal{R} \lambda)_i &= 1 \text{ mod } 2 \end{aligned} \nn\ee
These are respectively the statements that $(-1)^{F_L}$ and $(-1)^{F_R}$ lie within the $U(1)^N_\mathcal{R}$ group. For example, if $\lambda$ is a solution to the first condition, then $(-1)^{F_L}$ is given by the $U(1)^N_\mathcal{R}$ symmetry transformation with parameter $x = \lambda / 2$.

\subsubsection*{Recovering the $\mathbb{Z}_8$ Index}

Using the first of the conditions above, we can easily show that a boundary state preserving $\mathbb{Z}_2 \times \mathbb{Z}_2$ can only exist when the number of Dirac fermions $N$ is a multiple of 4. Since each Dirac fermion comprises two Majorana fermions, we recover the result stated in the introduction that the number of Majorana fermions must be a multiple of 8.

\para
For the proof, we need only to examine the length-squared of $\lambda$. Since $\mathcal{R}$ is orthogonal, we have
\be \lambda^2 = (\mathcal{R} \lambda)^2 \nn\ee
The vector $\lambda$ on the left is an $N$-component vector with odd components, so its length-squared is $N$ mod 8. Meanwhile, on the right hand side, $\mathcal{R} \lambda$ has even components, hence has length-squared 0 mod 4. Equating the two, we learn that
\be N = 0 \text{ mod } 4 \nn\ee
as claimed.

\subsection{Some Examples}

We now turn to some examples. We start by constructing all $\mathbb{Z}_2 \times \mathbb{Z}_2$ preserving states with $N=4$ Dirac fermions. We then briefly discuss the most stable such states with higher $N$.

\subsection*{The Minimal Case of 4 Dirac Fermions}

For the case of $N=4$ Dirac fermions, a parameterisation of the chiral-parity preserving boundary states can be given using the results of \cite{baake, loquias}.

\para
First, we briefly recount how a $4 \times 4$ rational orthogonal matrix can be parametrised in terms of a pair of integer quaternions, based on the isomorphism $SO(4) = (SU(2) \times SU(2)) / \mathbb{Z}_2$, following \cite{baake}. To each pair of integer quaternions $(\mathbf{p}, \mathbf{q})$, we associate an $SO(4)$ matrix $\mathcal{R}$, implicitly defined by
\be \mathcal{R} \mathbf{x} = \frac{1}{|\mathbf{p} \mathbf{q}|} \; \mathbf{q} \, \mathbf{x} \, \bar{\mathbf{p}} \nn\ee
where on the left $\mathbf{x}$ is regarded as a 4-component vector, acted on by the matrix $\mathcal{R}$, while on the right it is regarded as a quaternion. In order that $\mathcal{R}$ be rational, we must impose the constraint $|\mathbf{p} \mathbf{q}| \in \mathbb{Z}$. Without loss of generality, we may also assume that $\mathbf{p}$ and $\mathbf{q}$ are \emph{primitive}, meaning their four components are coprime. Then the set of all such pairs $(\mathbf{p}, \mathbf{q})$ gives a parametrisation of all $4 \times 4$ rational, special-orthogonal matrices $\mathcal{R}$, uniquely up to an overall sign redundancy $(\mathbf{p}, \mathbf{q}) \rightarrow (-\mathbf{p}, -\mathbf{q})$. (The remaining matrices with $\det(\mathcal{R}) = -1$ can be parametrised in an identical way, simply by multiplying the above expression by a reflection in the first coordinate.)

\para
Next, we turn to the conditions on $\mathbf{p}$ and $\mathbf{q}$ for the $\mathbb{Z}_2 \times \mathbb{Z}_2$ symmetry to be preserved. This question is addressed in \cite{loquias}, but to connect with their work we need to recast the condition for $\mathbb{Z}_2 \times \mathbb{Z}_2$ invariance in a slightly different form. It is straightforward to show that the following ratio of indexes can only take on the two possible values
\be \frac{[\mathbb{Z}^N : \mathbb{Z}^N \cap \mathcal{R}^{-1} \mathbb{Z}^N]}{[D_N : D_N \cap \mathcal{R}^{-1} D_N]} = 1 \text{ or } 2 \nn\ee
and that the value $2$ is attained precisely when the $\mathbb{Z}_2 \times \mathbb{Z}_2$ symmetry is preserved. Proposition~11 of \cite{loquias} then tells us that this happens if and only if
\begin{itemize}
\item Exactly one of $|\mathbf{p}|^2$, $|\mathbf{q}|^2$ is a multiple of 4.
\item Both $|\mathbf{p}|^2$ and $|\mathbf{q}|^2$ are 0 mod 4, and $\mathbf{p} \cdot \mathbf{q} \neq 0$ mod 4.
\item Both $|\mathbf{p}|^2$ and $|\mathbf{q}|^2$ are 2 mod 4, and $\mathbf{p} \cdot \mathbf{q}$ is odd.
\end{itemize}

\para
In Section~\ref{contsecproofexamples}, we first met the 4-fermion dyon state, whose importance was first emphasised by Maldacena and Ludwig \cite{juan} as an example of a particularly symmetric boundary state. It is also the simplest example of a $\mathbb{Z}_2 \times \mathbb{Z}_2$-preserving boundary state, a connection that was first made in \cite{shin1}. To see how it sits within the above parametrisation, we choose $p = (1, 1, 1, 1)$ and $q = (1, -1, -1, -1)$, and apply a reflection in the first coordinate to account for the fact that the dyon has determinant $-1$. This then gives us back the matrix $\mathcal{R}$ for the dyon that we met earlier,
\be \text{dyon}_4 \; = \; \frac{1}{2} \begin{pmatrix} +1 & -1 & -1 & -1 \\ -1 & +1 & -1 & -1 \\ -1 & -1 & +1 & -1 \\ -1 & -1 & -1 & +1 \end{pmatrix} \nn\ee
The dyon is actually the simplest of an infinite tower of $\mathbb{Z}_2 \times \mathbb{Z}_2$-preserving boundary states, in the sense that it has the lowest Affleck-Ludwig central charge $g_\mathcal{R}$. The full spectrum of allowed $g_\mathcal{R}$ values of all such states, together with the number of matrices supporting each one, reads:
\begin{center}
  \begin{tabular}{l|lllllllllll}
    $(g_\mathcal{R})^2$ & 2 & 6 & 10 & 14 & 18 & 22 & 26 & 30 & 34 & 38 & \dots \\
    \hline
    $ \# \mathcal{R} \, / \, 2^5 4!$ & 1 & 16 & 36 & 64 & 168 & 144 & 196 & 576 & 324 & 400 & \dots
  \end{tabular}
\end{center}
The first entry corresponds to the venerable dyon state. Along the top, we see that the possible $(g_\mathcal{R})^2$ values are $2 + 4k$ for $k \geq 0$, with their multiplicities in the bottom row given by sequence A031360 in OEIS \cite{oeis}. In particular this sequence never vanishes, showing that all values $(g_\mathcal{R})^2 = 2 + 4k$ are actually attained.

\subsubsection*{The Stablest Boundary State}

We have seen that $N$ Dirac fermions admit a $\mathbb{Z}_2 \times \mathbb{Z}_2$ preserving boundary only when $N$ is a multiple of 4, and in the minimal case $N = 4$, the constraint of symmetry invariance restricts the allowed spectrum of central charges $g_\mathcal{R}$, with the $\text{dyon}_4$ being the simplest. In this final section, we explore how this story generalises for higher $N$.

\para
For a larger number $N = 4k$ of Dirac fermions, our options for boundary states are considerably increased. For example, it is always possible simply to group the fermions into bunches of 4, and write down the dyon state for each one. This corresponds to choosing the matrix
\be \mathcal{R} = \underbrace{\text{dyon}_4 \oplus \dots \oplus \text{dyon}_4}_k \quad \text{with} \quad (g_\mathcal{R})^2 = 2^k \nn\ee
On the other hand, we also have the option of writing down a dyon state for all $4k$ fermions. This is also a $\mathbb{Z}_2 \times \mathbb{Z}_2$ preserving boundary state\footnote{This is because $\LambdaR$ consists of all integer vectors with component-sum a multiple of $2k$. As $2k$ is even, the boundary state satisfies the condition for $\mathbb{Z}_2 \times \mathbb{Z}_2$ invariance.}, and has central charge
\be \mathcal{R} = \text{dyon}_{4k} \quad \text{with} \quad (g_\mathcal{R})^2 = 2k \nn\ee
For $k \geq 3$, we see that the irreducible dyon has lower central charge than the first state. (For $k = 1,2$ they tie.) What's more, we conjecture that the dyon actually has the lowest central charge among \emph{all} $\mathbb{Z}_2 \times \mathbb{Z}_2$ preserving boundary states. Under the correspondence between boundary states and gapped phases, it would then be the `most generic' gapped phase preserving the $\mathbb{Z}_2 \times \mathbb{Z}_2$ symmetry, in the sense of having no relevant deformations that might destabilise it to a simpler phase corresponding to a boundary state with a smaller $g_\mathcal{R}$.

\para
The evidence for this conjecture is based on a randomised exploration of the space of $\mathbb{Z}_2 \times \mathbb{Z}_2$-preserving boundary states. To generate all such states, we use a modification of the null-vector construction \cite{haldane, juven}. In the unmodified null-vector construction, one constructs the charges $Q_{\alpha i}$ and $\bar{Q}_{\alpha i}$ iteratively, one $\alpha$ at a time, at each step picking the charges to ensure that $U(1)_\alpha$ has no anomaly either with itself or with $U(1)_\beta$ for $\beta < \alpha$. To modify this construction to produce precisely the chiral-parity-preserving states, our earlier results show that we need only ensure the charges of the first symmetry $\alpha = 1$ obey
\be Q_{1 i} = \text{odd} \qquad \bar{Q}_{1 i} = \text{even} \nn\ee
Choosing the charges of the remaining symmetries as before and setting $\mathcal{R} = \bar{Q}^{-1} Q$ then gives the most general chiral-parity-preserving state.

\para
With this recipe in place, we are finally in a position to explore the space of $\mathbb{Z}_2 \times \mathbb{Z}_2$ preserving boundary states. Combining the results for higher $N$ with those we've already seen for $N = 4$, we find the following:
\begin{itemize}

\item When $N = 4$ or $N = 8,16,24,\dots$, the possible central charges are
\be (g_\mathcal{R})^2 \in \{ \, N/2 + 4r : r \geq 0 \, \} \nn\ee

\item When $N = 12,20,28,\dots$, the possible central charges are
\be (g_\mathcal{R})^2 \in \{ \, N/2 + 2r : r \geq 0 \, \} \nn\ee

\end{itemize}
In both cases, the minimum value attained is $N/2$. This is the central charge of the dyon, which confirms its role as the simplest chiral-parity-preserving boundary state.


\begin{thebibliography}{1}

\bibitem{fk}   L.~Fidkowski and A.~Kitaev,
  ``{\it The effects of interactions on the topological classification of free fermion systems},''
  Phys.\ Rev.\ B {\bf 81}, 134509 (2010)
  [arXiv:0904.2197 [cond-mat.str-el]].
  
\bibitem{ryugso} S. Ryu and  S-C. Zhang, ``{\it  Interacting topological phases and modular invariance}", 
Phys.\ Rev.\ B {\bf 85}, 245132 (2012), [arXiv:1202.4484 [cond-mat.str-el]].  

\bibitem{qi} X-L Qi,  ``{\it A new class of (2+1)-d topological superconductor with ${\bf Z}_8$ topological classification}"
New J. Phys. 15 (2013) 065002, [arXiv:1202.3983 [cond-mat.str-el]]



\bibitem{ryu2} H. Yao and  S. Ryu  ``{\it Interaction effect on topological classification of superconductors in two dimensions}", Phys. Rev. B 88, 064507 (2013), [arXiv:1202.5805 [cond-mat.str-el]] 



\bibitem{bentov}   Y.~BenTov,
  ``{\it Fermion masses without symmetry breaking in two spacetime dimensions},''
  JHEP {\bf 1507}, 034 (2015)
  [arXiv:1412.0154 [cond-mat.str-el]].
  
\bibitem{carl}  D.~Tong and C.~Turner,
  ``{\it Notes on 8 Majorana Fermions,}''
  SciPost Phys. Lect. Notes 14 (2020)
  [arXiv:1906.07199 [hep-th]].  
  
  
    \bibitem{shin1}   G.~Y.~Cho, K.~Shiozaki, S.~Ryu and A.~W.~W.~Ludwig,
  ``{\it Relationship between Symmetry Protected Topological Phases and Boundary Conformal Field Theories via the Entanglement Spectrum},''
  J.\ Phys.\ A {\bf 50}, no. 30, 304002 (2017)
  [arXiv:1606.06402 [cond-mat.str-el]].
  
  
\bibitem{shin2}   B.~Han, A.~Tiwari, C.~T.~Hsieh and S.~Ryu,
  ``{\it Boundary conformal field theory and symmetry protected topological phases in $2+1$ dimensions},''
  Phys.\ Rev.\ B {\bf 96}, no. 12, 125105 (2017)
  [arXiv:1704.01193 [cond-mat.str-el]].
 
 \bibitem{ep}  E.~Eichten and J.~Preskill,
  ``{\it Chiral Gauge Theories on the Lattice},''
  Nucl.\ Phys.\ B {\bf 268}, 179 (1986).
  
  \bibitem{juven}   J.~Wang and X.~G.~Wen,
  ``{\it Non-Perturbative Regularization of 1+1D Anomaly-Free Chiral Fermions and Bosons: On the equivalence of anomaly matching conditions and boundary gapping rules},''
  arXiv:1307.7480 [hep-lat].
 
 
  \bibitem{slagle} K.~Slagle, Y.~Z.~You and C.~Xu,
  ``{\it Exotic quantum phase transitions of strongly interacting topological insulators},''
  Phys.\ Rev.\ B {\bf 91}, no. 11, 115121 (2015)
  [arXiv:1409.7401 [cond-mat.str-el]].

\bibitem{simon} 
  S.~Catterall,
  ``{\it Fermion mass without symmetry breaking},''
  JHEP {\bf 1601}, 121 (2016)
  [arXiv:1510.04153 [hep-lat]].
 	
\bibitem{ayyar}   V.~Ayyar and S.~Chandrasekharan,
  ``{\it Origin of fermion masses without spontaneous symmetry breaking},''
  Phys.\ Rev.\ D {\bf 93}, no. 8, 081701 (2016)
  [arXiv:1511.09071 [hep-lat]].

\bibitem{you1}   Y.~Z.~You, Y.~C.~He, C.~Xu and A.~Vishwanath,
  ``{\it Symmetric Fermion Mass Generation as Deconfined Quantum Criticality},''
  Phys.\ Rev.\ X {\bf 8}, no. 1, 011026 (2018)
  [arXiv:1705.09313 [cond-mat.str-el]].

\bibitem{youtoo}   Y.~Z.~You, Y.~C.~He, A.~Vishwanath and C.~Xu,
  ``{\it From Bosonic Topological Transition to Symmetric Fermion Mass Generation},''
  Phys.\ Rev.\ B {\bf 97}, no. 12, 125112 (2018)
  [arXiv:1711.00863 [cond-mat.str-el]].

\bibitem{juven2}  J.~Wang and X.~G.~Wen,
  ``{\it A Solution to the 1+1D Gauged Chiral Fermion Problem},''
  Phys.\ Rev.\ D {\bf 99}, no. 11, 111501 (2019)
  [arXiv:1807.05998 [hep-lat]].
  
\bibitem{john}   J.~L.~Cardy,
  ``{\it Boundary Conditions, Fusion Rules and the Verlinde Formula},''
  Nucl.\ Phys.\ B {\bf 324}, 581 (1989).

   \bibitem{afflud}   I.~Affleck and A.~W.~W.~Ludwig,
  ``{\it Universal noninteger `ground state degeneracy' in critical quantum systems},''
  Phys.\ Rev.\ Lett.\  {\bf 67}, 161 (1991).
  
  \bibitem{sagi}   I.~Affleck and J.~Sagi,
  ``{\it Monopole catalyzed baryon decay: A Boundary conformal field theory approach},''
  Nucl.\ Phys.\ B {\bf 417}, 374 (1994)
  [hep-th/9311056].

\bibitem{juan}  J.~M.~Maldacena and A.~W.~W.~Ludwig,
  ``{\it Majorana fermions, exact mapping between quantum impurity fixed points with four bulk fermion species, and solution of the `unitarity puzzle'},''
  Nucl.\ Phys.\ B {\bf 506}, 565 (1997)
  [cond-mat/9502109].
  
        
\bibitem{us}  P.~Boyle~Smith and D.~Tong,
  ``{\it Boundary States for Chiral Symmetries in Two Dimensions},''
  arXiv:1912.01602 [hep-th].

\bibitem{us2} P.~Boyle~Smith and D.~Tong,
  ``{\it Boundary RG Flows for Fermions and the Mod 2 Anomaly}", 
    arXiv:2005.11314 [hep-th].

\bibitem{baake}M.~Baake, ``Solution of the coincidence problem in dimensions $d \leq 4$'', arXiv:math/0605222 [math.MG].


\bibitem{bachas}   C.~Bachas, I.~Brunner and D.~Roggenkamp,
  ``{\it A worldsheet extension of $O(d,d:Z)$},''
  JHEP {\bf 1210}, 039 (2012)
  [arXiv:1205.4647 [hep-th]].




\bibitem{haldane}F.~D.~M.~Haldane, ``Stability of Chiral Luttinger Liquids and Abelian Quantum Hall States'', arXiv:cond-mat/9501007.


\bibitem{loquias}M.~Loquias, P.~Zeiner, ``Coincidence indices of sublattices and coincidences of colorings'', arXiv:1506.00028 [math.MG].

\bibitem{oeis} The Online Encyclopedia of Integer Sequences \href{https://oeis.org/A031360}{https://oeis.org/A031360}
\end{thebibliography}
\end{document}